\documentclass[%
 reprint,
 amsmath,amssymb,
 aps,
]{revtex4-2}

\usepackage{graphicx}
\usepackage{dcolumn}
\usepackage{color}
\usepackage{bm}
\usepackage{amsmath}
\usepackage{amssymb}
\usepackage{epsfig}
\usepackage{amsfonts}
\usepackage{lineno,hyperref}
\usepackage{array}
\usepackage{microtype}
\usepackage{float}
\usepackage{multirow}
\usepackage{adjustbox}
\usepackage[english]{babel}
\usepackage{blindtext}
\usepackage[a4paper, total={7.5in, 10in}]{geometry}

\def \a{\alpha}

\def \l{\lambda}

\def \g{\gamma}

\def \be{\begin{equation}}
\def \ee{\end{equation}}
\def \ben{\begin{eqnarray}}
\def \een{\end{eqnarray}}
\def \o{\omega}
\def \O{\Omega}

\def \p{\partial}
\def \e{\epsilon}

\def \t{\theta}

\def \r{\rho}
\def \k{\kappa}
\def \R{\bar{R}}
\def \T{\bar{T}}
\def\be{\begin{equation}}
\def\ee{\end{equation}}
\def\bse{\begin{subequations}}
\def\ese{\end{subequations}}
\def\ba{\begin{array}}
\def\ea{\end{array}}
\def\bea{\begin{eqnarray}}
\def\eea{\end{eqnarray}}

\begin{document}

\title{Evaporation of dynamical horizon with the Hawking temparature in the {\bf K-}essence emergent Vaidya spacetime}

\author{Bivash Majumder}
\email{bivashmajumder@gmail.com} 
\affiliation{Department of Mathematics, Prabhat Kumar College, Contai, Purba Medinipur 721404, West Bengal, India}
 
\author{Saibal Ray}
\altaffiliation{saibal.ray@gla.ac.in}
\affiliation{Centre for Cosmology, Astrophysics and Space Science (CCASS), GLA University, Mathura 281406, Uttar Pradesh, India}
  
\author{Goutam Manna$^{a}$ }
\altaffiliation{goutammanna.pkc@gmail.com\\$^{a}$Corresponding author}
\affiliation{Department of Physics, Prabhat Kumar College, \\Contai, Purba Medinipur 721404, West Bengal, India}

\date{\today}

\begin{abstract}
In the {\bf K-}essence Vaidya Schwarzschild spacetime, we apply the dynamical horizon equation to measure the mass-loss due to Hawking radiation and the tunneling formalism (Hamilton-Jacobi method) to calculate the hawking temperature. Assuming the Dirac-Born-Infeld kind of non-standard action for the {\bf K-}essence here, the background physical spacetime is a static spherically symmetric black hole, and we constrain the {\bf K-}essence scalar field to be a function only of either forward or backward time. The {\bf K-}essence emergent gravity and the generalizations of Vaidya spacetime have been linked by Manna et al. In this paper, we use Sawayama's modified description of the dynamical horizon to show that the obtained findings deviate from the standard Vaidya spacetime geometry.
\end{abstract}


\maketitle

\section{Introduction}
The type Ia Supernova (SNe Ia), baryon acoustic oscillations (BAO), cosmic microwave background (WMAP7), and Planck findings ~\cite{Riess,Perlmutter,Komatsu,planck1} all clearly suggest that the late time universe is speeding and dominated by dark energy~\cite{Bahcall1,peebles1}, which indirectly contradicted traditional gravity theories. Several scientists have already started exploring various aspects of gravitational theories and come up with solutions that show cosmic acceleration. There are several theoretical ideas that have been thoroughly investigated. 

Many of the existing models of dark energy struggle because they rely on a very precise tuning of the initial energy density of the highest order. To avoid the necessity for such precise modifications, a non-canonical theory, a new family of scalar field models known as {\bf K-}essence theory, has been developed~\cite{Armendariz1,babi1,babi2,babi3,babi4,babi5,scherrer1,scherrer2}, in which the negative pressure originates from the scalar field's non-linear kinetic energy term

Although the magnitude of the energy decreases by many orders and remains constant in the dust-dominated period, the {\bf K-}essence theory is unable to remove the dustlike equation of state for a couple of dynamical reasons. But after a long time (about now), the field prevails over the matter density and leads the cosmos towards the cosmic acceleration. According to the {\bf K-}essence theory, kinetic energy dominates over potential energy of the {\bf K-}essence scalar field. Lorentz invariance is broken spontaneously by non-trivial dynamical solutions of the {\bf K-}essence equation of motion with non-canonical kinetic terms. Furthermore, it modifies the metric for the perturbations close to these solutions. In the curved spacetime referred to as {\it emergent or analogue}, the perturbation propagates along with the metric. Taking the Dirac-Born-Infeld (DBI) model ~\cite{born1,born2,born3} into account, Manna et al.~\cite{gm3,gm4,gm5,gm6} have generated the simplest version of the emergent gravity metric $\bar G_{\mu\nu}$. It's important to note that this form is not conformally equal to the standard gravitational metric $g_{\mu\nu}$. A form of the Lagrangian for the {\bf K-}essence model ~\cite{Armendariz1,babi1,babi2,babi3,babi4,babi5,scherrer1,scherrer2} is $L=-V(\phi)F(X)$ where $X=\frac{1}{2}g^{\mu\nu}\nabla_{\mu}\phi\nabla_{\nu}\phi$.

There exists another form of Lagrangian~\cite{Tian} such as $L=[1+f(y)]X+[1+g(y)]V_{exp}$, where $V_{exp}=V_{0}~exp(-\lambda\phi)$, $V_{0}$ and $\lambda$ are constants, $y=X/V_{exp}$, and $f(y)$ and $g(y)$ are arbitrary functions. The authors of this paper use the {\bf K-}essence model to the study of primordial dark energy. In general, the Lagrangian can depend on arbitrary functions of $\phi$ and $X$. This theory is unique since it only accounts for the radiation epoch when tracing the background energy density. The {\bf K-}essence theory is more applicable now than ever before since the matter density in the expanding cosmos is falling faster than the energy density. The dark energy component of the {\bf K-}essence models, in which the speed of sound does not exceed the speed of light, may be able to account for the variations in the cosmic microwave background (CMB) at large angular scales~\cite{Erickson,DeDeo,Bean}. On the other hand, the {\bf K-}essence theory may be employed just from a gravitational or geometrical aspect~\cite{gm1,gm2,Ray} other than the dark energy model, because the existence of dark energy is still debatable~\cite{Nielsen} according to current observation~\cite{Ade}.

In addition, we analyze the non-canonical Lagrangian from a different perspective. In general, the canonical or standard form of the Lagrangian is $L=T-V$, where $T$ is the kinetic energy and $V$ is the potential energy of the system. However, according to Goldstein and Rana~\cite{Goldstein,Rana}, the general non-canonical form of Lagrangian leads to the canonical form given a particular condition. Because forces in scleronomic systems cannot be generated from any potential, the canonical Lagrangian has no explicit time dependency. Again, all scleronomic systems are not always conservative for systems exposed to dissipative processes. The canonical Lagrangian is easily derived from the non-canonical one. Fundamentally, $L$ does not have a unique functional form since the form of the Euler-Lagrange equations of motion may be kept for a variety of Lagrangian options ~\cite{Goldstein,Rana}. Furthermore, in special relativistic dynamics, the classical idea of $L(= T-V)$ is no longer appropriate ~\cite{Rana,Landau}. As a result, we may infer that the generic form of the Lagrangian is non-canonical type.

It is widely agreed that a black hole is the result of a gravitational collapse and emits thermal radiation as if it were very hot, with its temperature being directly proportionate to its surface gravity ~\cite{haw1,haw2,haw3,haw4,haw5,haw6,haw7,haw8,haw9,haw10}. Assuming the spacetime is to be static or stationary  ~\cite{haw1}, the black hole's mass will slowly diminish due to thermal radiation as long as the radiation emitted is negligible in comparison to the black hole's mass energy. It may be enhanced by using the Einstein equation for sufficiently large radiation. In this connection, a non-static solution of the Einstein's field equations for spheres of fluids radiating energy has been derived by Vaidya~\cite{vai1,vai2}. He has also established the nonstatic analogs of Schwarzschild's interior solution in~\cite{vai3,vai4} and solved the problem of gravitational collapse with radiation in~\cite{vai5}. This solution satisfies the physical feature of allowing a positive definite value of the density of collapsing matter and ensures that, from the perspective of a stationary observer at infinity, the total luminosity of the object is zero as it collapses to the Schwarzschild's singularity. It makes sense to think of the Vaidya spacetime~\cite{vai1,vai2,vai3,vai4,vai5} as a non-stationary Schwarzschild spacetime. Husain~\cite{husain} and Wang et al.~\cite{wang} proposed the generalized Vaidya spacetime, which is analogous to the gravitational collapse of a null fluid. Recently, Manna et al.~\cite{gm1,gm2} has established the {\bf K-}essence generalizations of Vaidya spacetime, in which the time dependence of the metric is determined from the kinetic energy ($\phi_{v}^{2}$) of the {\bf K-}essence scalar field ($\phi$). 

In the following works~\cite{parikh,mitra1,mitra2,mitra3,mitra4}, authors have discussed about Hawking radiation ~\cite{haw1,haw2,haw3,haw4,haw5,haw6,haw7,haw8,haw9,haw10} using tunnelling mechanism and this process in semi-classical quantum mechanics  can be described by the method of complex path analysis which is proposed in ~\cite{padma1,padma2}. Kerner and Mann~\cite{mann} have established that the Hawking temperature is independent of the angular part of the spacetime in general. It is possible to use either the radial-geodesic or Hamilton-Jacobi methods to analyze Hawking radiation. The Hawking radiation of a static and stationary black hole was studied using the radial geodesic approach created by Parikh and Wilczek~\cite{parikh}. Since its initial popularity in the 1990s, the Hamilton-Jacobi method has been revived as a tool for investigating black hole's non-thermal radiation. By solving Hamilton-Jacobi equations, one may determine the particle activity of both stationary and non-stationary black holes using this approach. The Hawking radiation in both Vaidya and Vaidya-Bonnor spacetimes is investigated in~\cite{kuroda,siahaan,tang,wanglin,chen,jun,niu}. Considering the Vaidya-Bonnor-de Sitter black hole, Chen and Yang~\cite{chen} have fixed the total energy and charge while allowing those of the black hole to vary. According to their theory the particle crosses the horizon radially. Basically their study addresses Hawking radiation when the particle tunnels through the potential barrier and the black hole does not absorb or emit any additional particles at $\Delta T$. Since the particle penetrates through the barrier instantly, $\Delta T$ is infinitely tiny. This suggests that the black hole cannot absorb or emit particles at this moment.

The expression of the energy and angular momentum fluxes transported by gravitational waves across the dynamical horizons, as well as the equation describing the variation of the dynamical horizon radius, are obtained in ~\cite{ashtekar1,ashtekar2,ashtekar3,ashtekar4,hayward,badri}. The definition of {\it dynamical horizon} is: A smooth, three-dimensional, space-like submanifold $H$ in a space-time $\cal{M}$ is said to be a dynamical horizon if it can be foliated by a family of closed $2-$surfaces such that, on each leaf $S$, the expansion $\Theta_{(l)}$ of one null normal $l^{a}$ vanishes and the expansion $\Theta_{(n)}$ of the other null normal $n^{a}$ is strictly negative. However, the modified definition proposed by Sawayama~\cite{sawayama} is as follows:  A smooth, three-dimensional, spacelike or timelike submanifold $H$ in a space-time is said to be a dynamical horizon if it is foliated by a preferred family of 2-spheres such that, on each leaf $S$, the expansion $\Theta_{(l)}$  of a null normal $l^{a}$ vanishes and the expansion $\Theta_{(n)}$ of the other null normal $n^{a}$ is strictly negative. For an example~\cite{blau}: in Minkowski spacetime, (i) radially outgoing light-rays, $l =\partial_{v},~ v = t + r$, have expansion: $\Theta_{l}=\nabla_{\a}(\partial_{v})^{\a}=\frac{1}{r^2}\p_{\a}\Big(r^{2}(\p_{t}+\p_{r})^{\a}\Big)=+\frac{2}{r}>0$, and (ii) radially ingoing light-rays $n=\p_{u}, u = t-r$, have expansion: $\Theta_{n}=\nabla_{\a}(\partial_{u})^{\a}=\frac{1}{r^2}\p_{\a}\Big(r^{2}(\p_{t}-\p_{r})^{\a}\Big)=-\frac{2}{r}<0$, which are indicating that outgoing light-rays expand while ingoing light-rays contract. 

Following the work of Ashtekar and Galloway~\cite{ashtekar3}, under the concept of world tubes, if the marginally trapped tube (MTT) is \\
(i) spacelike, then it is called a dynamical horizon (DH) and under some conditions that it provides a quasi-local representation of an evolving black hole.
 
(ii) timelike, then the causal curves can transverse it in both inward and outward directions, where it does not represent the surface of a black hole in any useful sense, it is called a timelike membrane (TLM).

(iii) null, then it describes a quasi-local description of a black hole in equilibrium and is called an isolated horizon (IH).
 
It is now highly promising to determine the behavior of non-static mass under a massive radiation scenario in this non-canonical theory, for the reasons stated above. In this work, we use the dynamical horizon equation based on Sawayama~\cite{sawayama} and the tunneling formalism~\cite{parikh,mitra1,mitra2,mitra3,mitra4,padma1,padma2,kuroda,siahaan,gm3,gm4,gm5} to investigate the Hawking effect in the {\bf K-}essence emergent generalized Vaidya spacetime for purely gravitational view point.

The paper is structured as follows: Section II provides a short overview of the {\bf K-}essence emergent geometry and the corresponding {\bf K-}essence Vaidya spacetime. In Section III, we explain the dynamical horizons for the emergent Vaidya spacetime with the {\bf K-}essence using the Schwarzschild black hole as a background. The dynamical horizon equation for the {\bf K-}essence Vaidya Schwarzschild spacetime has been studied in depth in the Section IV. We have also presented in Section V the related Hawking radiation by discussing the dynamical horizon equation and the tunneling mechanism. The last Section VI serves as a summary of our findings.

\section{Brief review  of {\bf K}-essence and {\bf K}-essence-Vaidya geometry}

\subsection{{\bf K}-essence Geometry}
The Scalar field $\phi$ of {\bf K}-essence possesses action~\cite{babi1}-\cite{babi5}
\ben
S_{k}[\phi,g_{\mu\nu}]= \int d^{4}x {\sqrt -g} L(X,\phi),
\label{1}
\een
when it is minimally coupled to the background gravitational metric  $g_{\mu\nu}$ and where $X=\frac{1}{2} g^{\mu\nu}\nabla_{\mu}\phi\nabla_{\nu}\phi$ and the energy-momentum tensor is
\ben
T_{\mu\nu}\equiv \frac{2}{\sqrt {-g}}\frac{\delta S_{k}}{\delta g^{\mu\nu}}= L_{X}\nabla_{\mu}\phi\nabla_{\nu}\phi - g_{\mu\nu}L,
\label{2}
\een
where $L_{\mathrm X}= \frac{dL}{dX},~~ L_{\mathrm XX}= \frac{d^{2}L}{dX^{2}},
~~L_{\mathrm\phi}=\frac{dL}{d\phi}$ and $\nabla_{\mu}$ is the covariant derivative defined with respect to the gravitational metric $g_{\mu\nu}$.

The scalar field equation of motion is
\ben 
-\frac{1}{\sqrt {-g}}\frac{\delta S_{k}}{\delta \phi}= \tilde G^{\mu\nu}\nabla_{\mu}\nabla_{\nu}\phi +2XL_{X\phi}-L_{\phi}=0,
\label{3}
\een
where  
\ben
\tilde G^{\mu\nu}\equiv L_{X} g^{\mu\nu} + L_{XX} \nabla ^{\mu}\phi\nabla^{\nu}\phi
\label{4}
\een
and 
$1+\frac{2X  L_{XX}}{L_{X}} > 0$.

Using the conformal transformations
$G^{\mu\nu}\equiv \frac{c_{s}}{L_{X}^{2}}\tilde G^{\mu\nu}$ and $\bar G_{\mu\nu}\equiv \frac{c_{s}}{L_{X}}G_{\mu\nu}$, with
$c_s^{2}(X,\phi)\equiv{(1+2X\frac{L_{XX}}{L_{X}})^{-1}}$ we have \cite{gm3,gm4,gm5}
\ben \bar G_{\mu\nu}
=g_{\mu\nu}-\frac{L_{XX}}{L_{X}+2XL_{XX}}\nabla_{\mu}\phi\nabla_{\nu}\phi.
\label{5}
\een

In order to investigate the behavior of minor perturbations as they propagate in the preferred reference frame, when the background is at rest, Babichev et al. \cite{babi3} has proposed the definition of sound speed ($c_{s}$), as indicated above. The speed of sound cannot often be higher than the speed of light. However, there are times when field fluctuations may spread at a superluminal rate ($c_{s}>1$).

For Eqs. (1)--(4), to have any physical significance, $L_{X}\neq 0$ must always hold and $c_{s}^{2}$ must be positive definite.

Now the equation of motion (\ref{3}) simplifies to 
\ben 
-\frac{1}{\sqrt {-g}}\frac{\delta S_{k}}{\delta \phi}
= \bar G^{\mu\nu}\nabla_{\mu}\nabla_{\nu}\phi=0,
\label{6}
\een
if and only if $L$ is not an explicit function of $\phi$.

One may take a note that the emergent metric $\bar{G}_{\mu\nu}$ is not conformally equal to $g_{\mu\nu}$ for non-trivial spacetime configurations of $\phi$. The local causal structure of $\phi$ also differs from that specified by $g_{\mu\nu}$, indicating that $\phi$ possesses properties distinct from canonical scalar fields.

The Lagrangian of the DBI type~\cite{gm3,gm4,gm5,born1,born2,born3} is assumed to be
\ben
L(X,\phi)= 1-V(\phi)\sqrt{1-2X},
\label{7}
\een
for $V(\phi)=V=$constant~and~kinetic energy of~$\phi>>V$, i.e., $(\dot\phi)^{2}>>V$. However, when it comes to {\bf K-}essence fields, it is common for kinetic energy to be more dominant over potential. In such case, $c_{s}^{2}(X,\phi)=1-2X$. For scalar fields $\nabla_{\mu}\phi=\partial_{\mu}\phi$. Then Eq. (\ref{5}) becomes
\ben
\bar G_{\mu\nu}= g_{\mu\nu} - \partial _{\mu}\phi\partial_{\nu}\phi.
\label{8}
\een

In order to get to the {\bf K-}essence emergent gravity metric described in Eq. (\ref{8}), we first apply a conformal transformation to determine the identify of the inverse metric $G_{\mu\nu}$, and then perform a second conformal transformation to realize the mapping onto the metric given in Eq. (\ref{8}) for the Lagrangian (Eq. \ref{7})~\cite{gm3}.

The geodesic equation for the {\bf K-}essence theory in terms of the new Christoffel connections  $\bar\Gamma$ is \cite{gm3,gm4,gm5}
\ben
\frac {d^{2}x^{\alpha}}{d\l^{2}} +  \bar\Gamma ^{\alpha}_{\mu\nu}\frac {dx^{\mu}}{d\l}\frac {dx^{\nu}}{d\l}=0,
\label{9}
\een
where $\l$ is an affine parameter and  
\ben
\bar\Gamma ^{\alpha}_{\mu\nu} 
&=&\Gamma ^{\alpha}_{\mu\nu} -\frac {1}{2(1-2X)}\Big[\delta^{\alpha}_{\mu}\partial_{\nu}X
+ \delta^{\alpha}_{\mu}\partial_{\nu}X\Big].~~~~~~~~~~~
\label{10}
\een

It is worth noting that the symmetry of $\bar\Gamma$ is preserved by the interchange of $\mu$ and $\nu$ in the second term on the right-hand side of Eq. (\ref{10}). The second term is unique to the {\bf K-}essence Lagrangian and represents further interaction (forces). Also, note that the Einstein tensor is not the same as the bf K-essence emergent gravity metric ($\bar{G}_{\mu\nu}$)  (Eq. \ref{8}).

It follows that $\mathcal{G}_{\mu\nu}\equiv \bar{R}_{\mu\nu}-\frac{1}{2}\bar{G}_{\mu\nu}\bar{R}=\kappa \mathcal{T_{\mu\nu}}$ is the corresponding Emergent Einstein Equation, where $\mathcal{G}_{\mu\nu}$ is the emergent Einstein tensor, $\mathcal{T_{\mu\nu}}$ is the corresponding energy-momentum tensor, $\kappa=8\pi G$, $\R_{\mu\nu}$ is Ricci tensor and $\R~ (=\R_{\mu\nu}\bar{G}^{\mu\nu})$ is the Ricci scalar of the emergent spacetime. In this case, any emergent metric $\bar{G}_{\mu\nu}$ that satisfies the ``Emergent Einstein Equation" may be considered as a solution. Components of the emergent Einstein tensor ($\mathcal{G}_{\mu\nu}$), energy-momentum tensor ($\mathcal{T_{\mu\nu}}$), and energy conditions for a generalized {\bf K-}essence Vaidya metric are discussed in depth in \cite{gm1}. Notably, the authors of~\cite{gm7,gm8,gm9} have used the {\bf K-}essence emergent gravity metric (Eq. \ref{8}) or (Eq. \ref{12})(below) in cosmology, where they have assumed that the underlying metric is of the Friedmann-Lemaître-Robertson-Walker (FLRW) type.

\subsection{{\bf K}-essence-Vaidya Geometry}
The Eddington-Finkelstein line element for a general spherically symmetric static black hole~\cite{gm1}
\ben
ds^{2}=f(r)dv^{2}-2\e dvdr-r^{2}d\O^{2},
\label{11}
\een
where $d\O^{2}=d\t^{2}+\sin^{2}\t d\Phi^{2}$.

It is to note that when $\e = +1$, the Eddington advanced time (outgoing) is represented by the null coordinate $v$. Further, the Eddington retarded time (incoming) is represented by the null coordinate $v$ when $\e = -1$. 

From Eq. (\ref{8}) the emergent spacetime is described by the line element
\ben
dS^{2}=ds^{2}-\p_{\mu}\phi\p_{\nu}\phi dx^{\mu}dx^{\nu}.
\label{12}
\een

The emergent spacetime line element is 
\ben
dS^{2}=\Big[f(r)-\phi_{v}^{2}\Big]dv^{2}-2\e dvdr-r^{2}d\O^{2},
\label{13}
\een
if we assume the scalar field $\phi(x)=\phi(v)$ where $\phi_{v}=\frac{\p \phi}{\p v}$.

Note that, in general, spherical symmetry would only require that $\phi(x)=\phi(v,r)$, hence the assumption on $\phi$ contradicts local Lorentz invariance. However, the dynamical solutions break Lorentz invariance spontaneously in the {\bf K-}essence theory. Therefore, the {\bf K-}essence scalar field in the form we've chosen is physically acceptable in this context. It should also be noted that the Lorentz invariance of the Minkowski spacetime~\cite{babi3,babi4} is preserved by the action (Eq. \ref{1}), even if the dynamical solutions of the {\bf K-}essence scalar field violate it.

Now, we derive 
\ben
dS_V^2 = \left(1-\frac{2m(v,r)}{r} \right)dv^2 - 2dvdr -r^2 d\O^2,
\label{14} 
\een
by comparing the metric~\cite{husain,wang} of the generalized Vaidya spacetimes corresponding to the gravitational collapse of a null fluid (choose $\e=+1$) with the metric of the emergent spacetime (Eq. \ref{13}). where the mass function is
\ben
m(v,r) = \frac12 r\Big[1 + \phi_v^2 - f(r) \Big].
\label{15}
\een

For the generalized {\bf K-}essence emergent spacetime with the Vaidya spacetime provided $\phi_v \phi_{vv} > 0~;~1 +\phi_v^2 > f + rf_r~;~2f_r + rf_{rr} > 0 $ which have established by Manna et. al. \cite{gm1}, these forms of metrics (Eq. \ref{13}) or (Eq. \ref{14}) meet all the necessary energy conditions~\cite{haw-ellis} (weak, strong, dominating).

\section{Dynamical horizons}
Following the works of some authors~\cite{ashtekar2,sawayama}, we now discuss how the dynamical horizon of the {\bf K-}essence emergent Vaidya spacetime behaves.

When the kinetic energy of the {\bf K-}essence scalar field $\phi_{v}^{2}$ is present, the generalized Vaidya spacetime (Eq. \ref{13}) or (Eq. \ref{14}) may be expressed 
\ben
dS^{2}=F(v,r)dv^{2}-2dvdr-r^{2}d\O^{2},
\label{16}
\een
where $v^{a}=$ a null vector. 

Based on Sawayama's~\cite{sawayama} work, we can define 
\ben
a=\frac{dr}{dr^{*}},
\label{17}
\een
where $r^{*}$ is tortoise coordinate defined as $v=t+r^{*}$.

Thus the two null vectors are
\ben
l^{a}=\left[\begin{array}{c} l^{t}	\\
l^{r^{*}}\\
l^{\theta}	\\
l^{\Phi}
\end{array}\right]=
\left[\begin{array}{c} a^{-1}	\\
-a^{-1}\\
0	\\
0
\end{array}\right],
\label{18}
\een
associating to the null vector $v^{a}$ , and the other one is
\ben
n^{a}=\left[\begin{array}{c} n^{t}	\\
n^{r^{*}}\\
n^{\theta}	\\
n^{\Phi}
\end{array}\right]=
\left[\begin{array}{c} a^{-1}	\\
\frac{F}{F-2a} a^{-1}\\
0	\\
0
\end{array}\right].
\label{19}
\een

The expansions of the two null vectors $l^{a}$ and $n^{a}$, $\Theta_{(l)}$ and  $\Theta_{(n)}$ \cite{sawayama}, are  
\ben
\Theta_{(l)}=\frac{1}{r}(2F-a)
\label{20}
\een
and
\ben
\Theta_{(n)}=\frac{1}{ar}\left(\frac{-2F^{2}+aF-2a^{2}}{-F+2a}\right).
\label{21}
\een

One can note that as $\Theta_{(l)}=0 \Rightarrow 2F-a=0$, and the other null expansion $\Theta_{(n)}$ is strictly negative which are the required conditions for the dynamical horizon~\cite{ashtekar1,ashtekar2,sawayama}. As a result,  the horizons of our case are dynamical. The following concepts explain the aforementioned necessary requirements:\\

The intuition that a black hole emits nothing -- not even light -- leads to a strictly zero null expansion, and the fact that null matter disappears into a black hole leads to a strictly negative null expansion. Note that the dynamical horizon for the emergent Vaidya spacetime in terms of the {\bf K-}essence has been developed by Manna et. al.~\cite{gm1} on the basis of Ashtekar et al. ~\cite{ashtekar1,ashtekar2,ashtekar3,ashtekar4,hayward}.

\subsection{Schwarzschild black hole as background:} 
For Schwarzschild black hole as background, we consider $f(r)=1-\frac{2M}{r}$, then Eq. (\ref{16}) becomes
\ben
dS^{2}=\Big(1-\frac{2M}{r}-\phi_{v}^{2}\Big)dv^{2}-2dvdr-r^{2}d\O^{2},
\label{22}
\een
with
\ben
F=\Big(1-\frac{2M}{r}-\phi_{v}^{2}\Big)\equiv\Big(1-\frac{2m(v,r)}{r}\Big).
\label{23}
\een

The mass function can be provided as
\ben
m(v,r)=M+\frac{r}{2}\phi_{v}^{2},
\label{24}
\een
where the tortoise coordinate
\ben
r^{*}=\frac{1}{N}\Big[r+B~ln(r-B)\Big],
\label{25}
\een
with $N\equiv N(v)=(1-\phi_{v}^{2})$, $B=2M'(v)=\frac{2M}{N}$ and $M'(v)=\frac{M}{1-\phi_{v}^{2}}$. 

Please take note that in the aforementioned spacetime (Eq. \ref{22}), $\phi_{v}^{2}<1$ is always the case. The signature of this spacetime is ill-defined if $\phi_{v}^{2}>1$. Moreover, the {\bf K-}essence hypothesis is rendered useless if $\phi_{v}^{2}=0$. It follows that $\phi_{v}^{2}\neq 0$ holds good. Again, $\phi_{v}^{2}\neq 1$ since if $\phi_{v}^{2}=1$ in Eq. (\ref{22}), it does
not have a Newtonian limit \cite{Schutz}, which makes it unsuitable for describing astrophysical objects.
For these reasons, $\phi_{v}^{2}$ might be anywhere from $0$ and $1$. Importantly, this article assumes only that the background gravitational metric is Schwarzschild. Future research into the horizon's dynamical behavior may also take into consideration the general spherically symmetric background metrics. Additionally, it is noted that Manna et al. \cite{gm3,gm4} have shown that for a specific solution of the {\bf K-}essence scalar field, the emergent gravity metrics transfer onto the Barriola-Vilenkin (BV) and Robinson-Trautman (RT) type metric, respectively, when the Schwarzschild and Reissner-Nordstr{\"o}m metrics are considered as a backdrop. For instance, these BV and RT metrics are static, spherically symmetric emergent metrics.

In this case, we obtain 
\begin{align}
a&=F\Big[1-\frac{1}{N}\Big(\frac{dB}{dv}~ln(r-B)-\frac{B}{r-B}\frac{dB}{dv}\Big)\nonumber\\&
+\frac{1}{N^{2}}\Big(r+B~ln(r-B)\frac{dN}{dv}\Big)\Big],
\label{26}    
\end{align}
by taking the derivative of the preceding Eq. (\ref{25}) with respect to $r^{*}$. 

The dynamical horizon radius
\begin{align}
2F-a&=2F-F\Big[1-\frac{1}{N}\Big(\frac{dB}{dv}~ln(r-B)-\frac{B}{r-B}\frac{dB}{dv}\Big)\nonumber\\&+\frac{1}{N^{2}}\Big(r+B~ln(r-B)\frac{dN}{dv}\Big)\Big]=0,
\label{27}    
\end{align}
is found by solving $\Theta_{(l)}=0$. 

As a result, we have two possible solutions for $r$: 
\ben
r_{D}=2M'(v)=\frac{2M}{1-\phi_{v}^{2}}=\frac{2M}{N},
\label{28}
\een\
by solving $F=0$ and
\ben
(r_{D}-B)^{B}e^{r_{D}}=e^{-vN},
\label{29}
\een
by solving  
$1+\frac{d}{dv}\Big[\frac{B}{N}~ln(r_{D}-B)\Big]-\frac{r_{D}}{N^{2}}\frac{dN}{dv}=0$. 

The Wright Omega function ($\o$)~\cite{corless1} may be used to express this value of $r_{D}$ as 
\ben
r_{D}=B\Big[1+\o(Z)\Big]=\frac{2M}{N}\Big[1+\o(Z)\Big]
\label{30}
\een
with $Z=-[1+ln(B)+vC]$, $C=\frac{N^{2}}{2M}=\frac{(1-\phi_{v}^{2})^{2}}{2M}$ where the dynamical radius lies outside $r_{D}=\frac{2M}{N}$.

The Wright Omega function ($\o$) is a single-valued function, defined in terms of the Multi-valued Lambert W function \cite{corless2} as
$\o(Z)=W_{\mathcal{K}(Z)}(e^{Z})$ where $\mathcal{K}(Z)(=[\frac{(Im(Z)-\pi)}{2\pi}])$ is the unwinding number of $Z$. The sign of this unwinding number is such that $ln(e^{Z})=Z+2\pi i\mathcal{K}(Z)$ which is opposite to the sign used in \cite{corless3}. The algebraic properties \cite{corless1} of the Wright Omega function ($\o$) are
\begin{align}
\frac{d\o}{dZ}=\frac{\o}{1+\o},  \label{31}
\end{align}

\begin{align}
 \int{\o^{n}dZ}=
\Bigg\{
 \begin{split}\frac{\o^{n+1}-1}{n+1}+\frac{\o^{n}}{n}~~if~n\neq -1, \\
 ln~\o -\frac{1}{\o} ~~~~~~~~~if~n=-1,
\label{32} 
\end{split}
\end{align}
having the analytical property $Z=\o+ln~\o$.

Here Eq. (\ref{29}) allows us to re-establish Sawayama's~\cite{sawayama} finding 
\ben
r_{D}=2M(v)\left[1+e^{-v/2M(v)}\right],
\label{33}
\een
in the usual Vaidya spacetime without the {\bf K-}essence scalar field $\phi$ if we consider $m(v,r)\equiv M(v)$ and $\phi_{v}^{2}=0$.

\section{Dynamical horizon equation}
The Ricci scalar ($\R$) and Ricci tensors ($\R_{\mu\nu}$) of the {\bf K-}essence emergent Vaidya spacetime may be derived from Eq. (\ref{16}) as
\begin{align}
&\R_{vv}=\frac{1}{r}\p_{v}F-\frac{F}{2}\p_{r}^{2}F-\frac{F}{r}\p_{r}F~;~\nonumber\\
&\R_{rv}=\R_{vr}=\frac{1}{2}\p_{r}^{2}F+\frac{1}{r}\p_{r}F~:~ \R_{rr}=0~;~\nonumber\\
&\R_{\t\t}=F+r\p_{r}F-1~;~\R_{\Phi\Phi}=sin^{2}\t\R_{\t\t}~;~\nonumber\\
&\R=-[\p_{r}^{2}F+\frac{4}{r}\p_{r}F+\frac{2}{r^{2}}(F-1)].
\label{34}
\end{align}

The components of the energy momentum tensor for the {\bf K-}essence emergent Vaidya spacetime may be derived from the ``emergent" Einstein's equation $\R_{\mu\nu}-\frac{1}{2}\bar{G}_{\mu\nu}\R=8\pi  \T_{\mu\nu}$ using these values for the Ricci scalar and the Ricci tensors (Eq. \ref{34}) as
\begin{align}
8\pi\T_{vv}=&\frac{1}{r}\p_{v}F+\frac{1}{r}F\p_{r}F+\frac{F}{r^{2}}(F-1),
\label{35}\\
8\pi\T_{vr}=&-[\frac{1}{r}\p_{r}F+\frac{1}{r^{2}}(F-1)],\label{36}\\
8\pi\T_{rr}=&0,\label{37}\\
8\pi\T_{\t\t}=&-[\frac{r^{2}}{2}\p_{r}^{2}F+r\p_{r}F]~;~8\pi\T_{\Phi\Phi}=\sin^{2}\t \T_{\t\t}, \label{38}    
\end{align}
where the gravitational constant $G=1$. 

Now according to Eq. (\ref{22}), for the Schwarzschild background case, the energy-momentum tensor components are:
\begin{align}
8\pi\T_{vv}=&-\frac{2}{r^{2}}[\p_{v}m+F\p_{r}m] \nonumber\\
=&-\left[\frac{2\phi_{v}\phi_{vv}}{r}+\frac{\phi_{v}^{2}}{r^{2}}(1-\frac{2M}{r}-\phi_{v}^{2})\right], \label{39}\\
8\pi\T_{vr}=&\frac{2}{r^{2}}\p_{r}m=\frac{\phi_{v}^{2}}{r^{2}}~or~8\pi\T_{vr^{*}}=\frac{2a}{r^{2}}\p_{r}m=a\frac{\phi_{v}^{2}}{r^{2}}, \label{40}
\\
8\pi\T_{rr}=&8\pi\T_{r^{*}r^{*}}=0, \label{41}    
\end{align}
for the spherically symmetric {\bf K-}essence emergent Schwarzschild Vaidya spacetime. 

It is worth noting that in our previous work~\cite{gm1} we derived the energy requirements of the emergent Vaidya spacetime with {\bf K-}essence (Eq. \ref{22}) for the Schwarzschild background. The energy conditions are: 
$\g\geq0,~\r\geq0,~P\geq0~ (\g\neq0)$, which satisfies the weak and strong energy conditions and $\g\geq0,~\r\geq0,~P\geq0~ (\g\neq0)$, which satisfy the dominant energy condition provided $\phi_{v}\phi_{vv}>0$, where $\g=\frac{2\phi_v \phi_{vv} }{\k r};~\r=\frac{\phi_v^2}{\k r^{2}}~and~P=0$ with $\kappa=8\pi G$. The associated energy-momentum tensor is of the type-II class \cite{gm1,haw-ellis}, which is characterized by a double null vector. It is possible to integrate the dynamical horizon equation precisely given in~\cite{ashtekar1,ashtekar2,sawayama} by deriving the energy-momentum tensor $\T_{\hat{t}l}$, as shown in \cite{sawayama} as
\begin{widetext}
\begin{align}
\frac{1}{2G}(\mathcal{R}_{2}-\mathcal{R}_{1})=\int_{\Delta H}T_{ab}\hat{\tau}^{a}\xi_{(\mathcal{R})}^{b}d^{3}V+\frac{1}{16\pi G}\int_{\Delta H} N_{\mathcal{R}}[ 	\mid\sigma	\mid^{2}+2	\mid\zeta	\mid^{2}]d^{3}V,~~\label{42}    
\end{align}    
\end{widetext}
where $\mathcal{R}_{2}$, $\mathcal{R}_{1}$ are the radii of the dynamical horizon, $T_{ab}$ is
the stress-energy tensor, $\mid\sigma	\mid^{2}=\sigma_{ab}\sigma^{ab}$, $	\mid\zeta	\mid^{2}=\zeta_{a}\zeta^{a}$, 
$\sigma_{ab}$ is the shear, $\zeta^{a}=\tilde{q}^{ab}\hat{r}^{c}\nabla_{c}l_{b}$, with the
two-dimensional metric $\tilde{q}^{ab}$, and $\xi_{(\mathcal{R})}^{a}=N_{\mathcal{R}}l^{a}$ with $N_{\mathcal{R}}=	\mid\p \mathcal{R}	\mid$, where $\mathcal{R}=$radius of the dynamical horizon.

This is the dynamical horizon equation (\ref{42}), and it describes how the horizon radius varies as a function of matter flow, shear, and expansion. The right-hand side of the dynamical equation is simplified since the second component disappears when the system is spherically symmetric.

To begin with, we may express $\T_{tl}$ as the combination of $\T_{vv}$ and $\T_{vr^{*}}$, where 
\begin{align}
\T_{tl}=\T_{vv}-\T_{vr^{*}}=-\frac{1}{4\pi r^{2}}\frac{5}{2}\p_{v}m
=-\left(\frac{1}{4\pi r^{2}}\right)\frac{5}{2}r\phi_{v}\phi_{vv}. \label{43}    
\end{align}

Assuming $\hat{t}^{a}$ is the unit vector in the direction of $t^{a}$, we get
\ben
\T_{\hat{t}l}=-\frac{1}{4\pi r^{2}}\frac{5}{2}(\p_{v}m) F^{-1},
\label{44}
\een
with $F$ is defined in Eq. (\ref{23}). 

At the horizon  ${r=r_{D}}$, the expression of $\T_{\hat{t}l}$ is $\T_{\hat{t}l} =\frac{1}{4\pi r_{D}^{2}}\frac{5r_{D}}{4}(\p_{v}N)F^{-1}$ using $m\equiv m(v,r)=M+\frac{r}{2}(1-N)$. We therefore generate the following terms: 
\begin{widetext}
\begin{align}
\frac{dr_{D}}{dv}=\Bigg[-\frac{2M}{N^{2}}\Big(1+\o(Z)\Big)+\frac{2M}{N}~\frac{\o(Z)}{1+\o(Z)}\times\Big(\frac{1}{N} 
-\frac{Nv}{M}\Big)\Bigg]\p_{v}N  -\frac{N~\o(Z)}{1+\o(Z)},
 \label{45}    
\end{align} 
\end{widetext}
to assess the dynamical horizon integration (Eq. \ref{42}) in terms of the Wright $\o$ function using Eq. (\ref{30}). 

Using the above Eq. (\ref{45}), we have
\begin{widetext}
\ben
\p_{v}N=\frac{N~\o(Z)}{1+\o(Z)}~
\Bigg[-\frac{2M}{N^{2}}\Big(1+\o(Z)\Big)+\frac{2M}{N}~\frac{\o(Z)}{1+\o(Z)}~\Big(\frac{1}{N}-\frac{Nv}{M}\Big)\Bigg]^{-1},
\label{46}
\een
\end{widetext}
at the horizon. 

Again, from Eq. (\ref{30})
\ben
\frac{dr_{D}}{dN}=\Big(\frac{N}{\p_{v}N}\Big)\frac{\o(Z)}{1+\o(Z)}.
\label{47}
\een

Now, rewriting Eq. (\ref{23}) in terms of Write $\o$ function as  
\ben
F=\frac{N~\o(Z)}{1+\o(Z)}
\label{48} 
\een
and also from Eq. (\ref{24}), we have
\begin{align}
 \p_{v}m=-\Big(\frac{r_{D}}{2}\Big)\p_{v}N.
\label{49}   
\end{align}

Now, by changing the order of integration $r_{D}$ to $N$ in Eq. (\ref{42}), we get
\begin{align}
\frac{1}{2}\left[\frac{2M}{N}\big(1+\o(Z)\big)\right]_{N_{1}}^{N_{2}}=\frac{5}{4}\int_{N_{1}}^{N_{2}} \frac{2M}{N}\big(1+\o(Z)\big)dN,
\label{50}    
\end{align}
since the functions $\p_{v}N$ and $F^{-1}$ with fixed $r_{D}$ are used only in the integration. 

By substituting the aforementioned Eq. (\ref{50}) into Eq. (\ref{42}) as well as using Eqs. (\ref{44}) to (\ref{49}) and then considering the limit $N_{2}\rightarrow N_{1}=N$, one can obtain 
\begin{widetext}
\begin{align}
-\frac{M}{N^{2}}\Big(1+\o(Z)\Big)+\frac{M}{N}\frac{\o(Z)}{(1+\o(Z))}\Big(\frac{1}{N}-\frac{Nv}{M}\Big)-\frac{5}{2}\frac{M}{N}\Big(1+\o(Z)\Big)=0.
\label{51}    
\end{align}  
\end{widetext}

This is the dynamical horizon equation (\ref{51}) for the spherically symmetric {\bf K-}essence emergent Schwarzschild Vaidya spacetime. When the kinetic energy of the {\bf K-}essence scalar field is present, the usual Vaidya dynamical horizon equation in~\cite{sawayama} is radically changed.

\section{Hawking Radiation in the {\bf K-}essence Schwarzschild-Vaidya spacetime}
Now we are going to discuss about the {\it Hawking radiation}~ \cite{haw1,haw2,haw3,haw4,haw5,haw6,haw7,haw8,haw9,haw10}. In order to find a solution, we investigate two different approaches: (1) the dynamical horizon equation (\ref{42}) and (2) the {\bf K-}essence Schwarzschild-Vaidya metric (Eq. \ref{22}) using a tunneling mechanism~\cite{parikh,mitra1,mitra2,mitra3,mitra4,padma1,padma2,kuroda,siahaan}.

\subsection{Dynamical horizon equation}

Taking into account Candelas's conclusion \cite{cand} for matters on the dynamical horizon, which is appropriate close to the horizon \cite{sawayama} and supposes that spacetime is almost static, from Eq. (\ref{22}) and $r\sim 2m (\equiv \frac{2M}{N})$, we get
\begin{align}
\T_{tl}&=\frac{-1}{2\pi^{2}(1-\frac{2m}{r})}\int_{0}^{\infty}\frac{w^{3}dw}{e^{8\pi mw}-1}\nonumber\\
&=\frac{-1}{2m^{4}\pi^{2}c(1-\frac{2m}{r})},
\label{52}    
\end{align}
with $c=15\times 8^{4}=61440$ and $\int_{0}^{\infty}\frac{w^{3}dw}{e^{bw}-1}=\frac{\pi^{4}}{15b^{4}}$. 

In the dynamical horizon equation, the matter-energy of Eq. (\ref{52}) is negative around $r\sim 2m$. Hence the {\bf K-}essence Schwarzschild-Vaidya black hole absorbs negative energy, causing the black hole's radius to shrink. At this juncture, one may note that the Schwarzschild mass is less than the {\bf K-}essence Schwarzschild-Vaidya mass as $\phi_{v}^{2}<1$, whereas larger at the horizon, where the {\bf K-}essence Schwarzschild mass $m(\equiv \frac{M}{1-\phi_{v}^{2}})$.
Also, the kinetic energy ($\phi_{v}^{2}$) of the {\bf K-}essence scalar field, as described by Eq. (\ref{24}), carries the dynamic behavior of the mass function $m(v,r)$.

Here, we employ the dynamical horizon equation in place of solving the entire Einstein equation with the backreaction, since we are only looking for information about matter close to the horizon. Following~\cite{sawayama}, we get
\begin{align}
\T_{\hat{t}l}=\frac{1}{2m^{4}\pi^{2}c(1-\frac{2m}{r})}.  \label{53}     
\end{align}

Again, one may obtain
\begin{widetext}
\begin{align}
\int_{r_{1}}^{r_{2}}4\pi r_{D}^{2}\T_{\hat{t}l}dr_{D}
&=b\int_{N_{1}}^{N_{2}}\Big[\frac{2M}{N} (1+\o(Z))\Big]^{2}\frac{(1+\o(Z))}{(N\o(Z))}\Big[M+\frac{M}{N}(1+\o(Z))(1-N)\Big]^{-4} \Big(\frac{dr_{D}}{dN}\Big) dN\nonumber\\
&=b\int_{N_{1}}^{N_{2}}\Big[\frac{2M}{N} (1+\o(Z))\Big]^{2}\Big[M+\frac{M}{N}(1+\o(Z))(1-N)\Big]^{-4}\Big(\frac{1}{\p_{v}N}\Big)  dN,
\label{54}    
\end{align}    
\end{widetext}
by changing the order of integration from $r_{D}$ to $N$ in the right hand side of the dynamical horizon equation (\ref{42}), where $b=\frac{2}{\pi c}=$ a constant and $\frac{dr_{D}}{dN}=\frac{N\o(Z)}{1+\o(Z)}\times\frac{1}{\p_{v}N}$.

Specifically, the {\bf K-}essence Schwarzschild-Vaidya metric has a time dependency that carries $\phi_{v}^{2}$, i.e., $N(=1-\phi_{v}^{2})$.

Here, we derive 
\begin{widetext}
\begin{align}
 \frac{1}{2}\Big[\frac{2M}{N}\Big(1+\o(Z)\Big)\Big]_{N_{1}}^{N_{2}}&=b\int_{N_{1}}^{N_{2}}\Big[\frac{2M}{N} \Big(1+\o(Z)\Big)\Big]^{2}\Big[M+\frac{M}{N}\Big(1+\o(Z)\Big)(1-N)\Big]^{-4}\Big(\frac{1}{\p_{v}N}\Big)dN\nonumber\\&+\frac{5}{4}\int_{N_{1}}^{N_{2}} \frac{2M}{N}\Big(1+\o(Z)\Big)dN,
\label{55}   
\end{align}  
\end{widetext}
by substituting Eq. (\ref{54}) into the right-hand side of the dynamical horizon Eq. (\ref{42}) and where $$r_{D}=B\Big[1+\o(Z)\Big]=\frac{2M}{N}\Big[1+\o(Z)\Big].$$ 

When we choose the limit $N_{2}\rightarrow N_{1}=N$ and $\o(Z)\neq 0$, the dynamical horizon Eq. (\ref{55}) simplifies to 
\begin{widetext}
\begin{align}
{2}\Bigg[1+\Big(1-\frac{1}{N}\Big)\Big(1+\o(Z)\Big)\Bigg]^{4}=\frac{4b(1+\o(Z))^{3}}{N^{3}\o(Z)}\Bigg[\frac{1}{2}-\frac{5(1+\o(Z))}{4}\Bigg(-\frac{1+\o(Z)}{N}+\frac{\o(Z)}{1+\o(Z)}\Big(\frac{1}{N}-\frac{Nv}{M}\Big)\Bigg)^{-1}\Bigg]^{-1},
\label{56}    
\end{align}  
\end{widetext}
as long as both $N$ and $M$ are non-zero. 

If $N$ is a function of $v$ according to the above Eq. (\ref{56}), then $\phi_{v}^{2}$ must also be a function of $v$ according to the same logic, because $\phi_{v}^{2}=1-N$. The Eq. (\ref{56}) contains the Wright omega function $\o(Z)$ (where $Z=-[1+ln(B)+vC]$) in a comprehensive fashion. As such, there is no method to derive $N$ analytically, making it quite distinct from Sawayama's~\cite{sawayama} transcendental equation (37).

In this subsection, however, we are primarily interested in determining the behavior of the black hole mass $m(v,r)$ with $v$ for fixed $M$. Here, we must first solve Eq. (\ref{56}) for $N$ for a given value of $M$ assuming $M\ne0$, while in ~\cite{sawayama}, the behavior of $M(v)$ may be found immediately using numerical techniques from Eq. (37) of Sawayama's work \cite{sawayama}. Then, by substituting into the formula $\phi_{v}^{2}=1-N$, we can determine what $\phi_{v}^{2}$ is when $M$ is a specific value. Last but not least, the relation (Eq. \ref{24}) allows us to observe the change in mass $m(v,r)$ as $v$ increases.

 \begin{figure*}
\includegraphics[width=14.0cm]{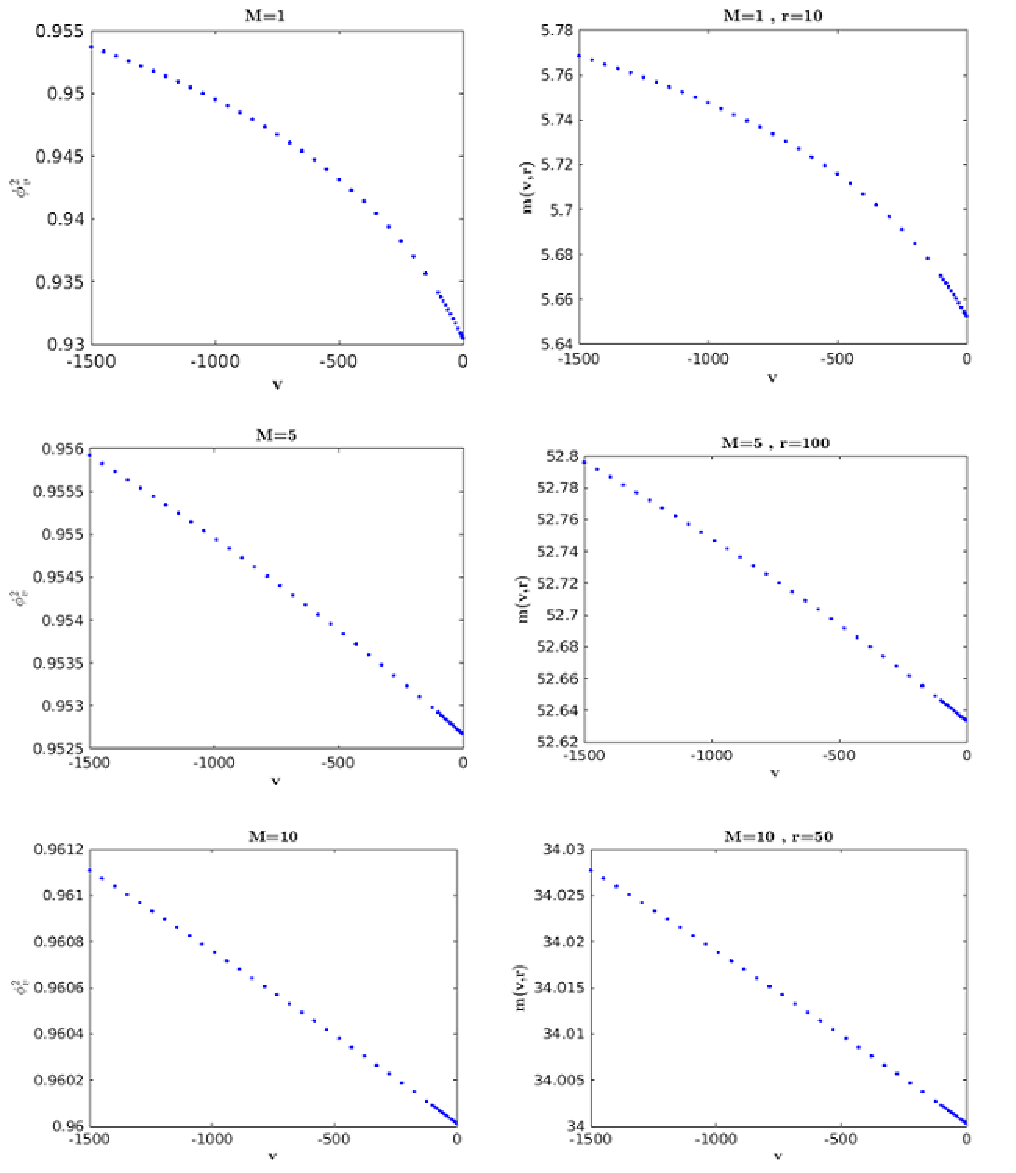}
 \caption{(Color online) Figures in the left column show the numerical solutions of $\phi^{2}_{v}$ for $M=1$, $M=5$, and $M=10$. Figures in the right column show the numerical solutions of $m(v,r)$ for $(M=1, r=10)$ , $(M=5, r=100)$ and $(M=10, r=50)$.}
 \label{fig:P}
\end{figure*}   

As shown in the left column of Fig. (\ref{fig:P}), if we numerically solve Eq. (\ref{56}) for $M=1$, $M=5$, and $M=10$, then the solutions of $\phi_{v}^{2}$ consistently decrease with increasing $v$. Now substituting these values of $\phi_{v}^{2}$ for $M=1$ and $r=10$ in Eq. (\ref{24}), we find that the black hole's mass $(m(v,r))$ decreases with increasing $v$ but does not go to zero, as seen from the first panel in the right column of Fig. (\ref{fig:P}). For $M=5$, $r=100$ and $M=10$, $r=50$, we find similar behavior of $m(v,r)$ when solving Eqs. (\ref{56}) by using Eq. (\ref{24}). These solutions are shown in the final two figures in the right column. In addition, it is said in \cite{sawayama} that the black hole's mass completely vanished at the moment when time ($v$) axis considered as $-\infty$ to zero (shown as $v=0$ in Fig. 2). In contrast, our black hole mass $(m(v,r))$ (Eq. \ref{24}) is not completely evaporated at $v=0$, even while taking into account the same kind of axis representation. In the {\bf K-}essence Schwarzschild-Vaidya spacetime, the blackhole mass diminishes but does not completely evaporate as shown by the decreasing behavior of $\phi_{v}^{2}$ and $m(v,r)$ with increasing $v$. 

Analytical proof of this situation may be derived from Eq. (\ref{24}): in the limiting case, $m(v,r)\rightarrow M$ as $\phi_{v}^{2}\rightarrow 0^{+}$ when $r$ and $M$ are fixed, or $m(v,r)\rightarrow M$ as $r\rightarrow 0$ when $\phi_{v}^{2}$ and $M$ are fixed since $M\ne 0$. This is in sharp contrast to Sawayama's \cite{sawayama} conclusion, according to which the black hole's mass always vanishes. However, if we write the mass $m(v,r)$ as $\frac{r}{2}\phi_{v}^{2}$ (using Eq. (\ref{24})) where $M$ is extremely tiny and negligible, then Eq. (\ref{22}) may be represented as (in the usual coordinate system): by rescaling, we may convert $dS^{2}=(1-\phi_{v}^{2})dt^{2}-\frac{dr^{2}}{(1-\phi_{v}^{2})}-r^{2}d\Omega^{2}$ to $dS^{2}=dt^{2}-dr^{2}-r^{2}(1-\phi_{v}^{2})d\Omega^{2}$. Because the re-scaled metric reflects a space with a deficit solid angle, the area of a sphere with radius $r$ is less than $4\pi r^{2}$, specifically, it is $(1-\phi_{v}^{2})4\pi r^{2}$ since $\phi_{v}^{2}$ takes the values in between $0$ and $1$. As a result, black holes of this type cannot totally evaporate since their associated spaces are not asymptotically flat but asymptotically bound.

\subsection{The tunneling formalism: Hamilton-Jacobi method}
Using the {\it tunneling approach}~\cite{parikh,mitra1,mitra2,mitra3,mitra4,padma1,padma2,kuroda,siahaan,gm3,gm4,gm5}, one may get the Hawking temperature for a massless particle in a black hole (Eq. \ref{22}) whose background is given by the Klein-Gordon equation 
\ben
\hbar^{2}\left(-\bar{G}\right)^{-1/2}\p_{\mu}\left(\bar{G}^{\mu\nu}\left(-\bar{G}\right)^{1/2}\p{\nu}\Psi\right)=0,
\label{57}
\een 
where $\Psi$ is in the form
\ben
\Psi=\textit{exp}\left(\frac{i}{\hbar}S+...\right).
\label{58}
\een

To find the leading order in $\hbar$ the Hamilton-Jacobi equation is 
\ben
\bar{G}^{\mu\nu}\p_{\mu}S\p_{\nu}S=0,
\label{59}
\een
where we consider $S$ is independent of $\theta$ and $\Phi$. 

Thus
\ben
2\p_{v}S\p_{r}S+\Big(1-\frac{2M}{r}-\phi_{v}^{2}\Big)\Big(\p_{r}S\Big)^{2}=0.
\label{60}
\een  

In the conventional Hamilton-Jacobi description, the action $S(v,r)$ may be broken down into two parts: the time component, denoted by $Ev$, and the radius component, denoted by $S_0(r)$, which normally only relies on radius. For black holes of varying masses, however, the energy of the departing particle must change over time since the metric coefficients rely on both radius and time. In view of this, the conventional method obviously fails. Yet, we may extend the procedure \cite{siahaan} if we want to determine the action $S(v,r)$. Hence, following~\cite{siahaan}, we can choose the action $S$ to be form 
\ben
S(v,r)=-\int_{0}^{v}E(v')dv'+S_{0}(v,r).
\label{61}
\een 

The term $\int_{0}^{v}E(v')dv'$ is simpler to understand due to the continuous and time-dependent energy of the discharged particles. So that \cite{siahaan}
\ben
\p_{v}S=-E(v)+\p_{v}S_{0}~~\text{and}~~\p_{r}S=\p_{r}S_{0}.
\label{62}
\een

Due to the fact that $S_{0}$ is proportional to $v$ and $r$, we have 
\ben
\frac{dS_{0}}{dr}=\p_{r}S_{0}+\frac{dv}{dr}\p_{v}S_{0}
=\p_{r}S_{0}+\frac{2}{F}\p_{v}S_{0},
\label{63}
\een
where we apply the formula $\frac{dv}{dr}=\frac{2}{F}$, $F$ being specified by the given Eq. (\ref{23}). 

The solution to (\ref{60}) is 
 \ben
 F\frac{dS_{0}}{dr}=2E(v),
 \label{64}
 \een
which has been obtained by plugging the results of (\ref{62}) and (\ref{63}) into (\ref{60}) and because $\p_{r}S_{0}\neq 0$. 

As a result, we may write the solution to $S_{0}$ as 
\ben
S_{0}=2E(v)\int \frac{dr}{F} \equiv 2E(v)\int \frac{dr}{(1-\frac{2M}{r}-\phi_{v}^{2})}\nonumber\\
=\frac{2E(v)}{N}\int\frac{r~dr}{r-2M/N}=2\pi i\frac{4ME(v)}{N^{2}},
\label{65}
\een
having $N=1-\phi_{v}^{2}$. 

Since $r$ is analytic within and on any simple closed contour $C$, considered in the positive sense and $\frac{2M}{N}$ is a point inside $C$, we have employed the Cauchy-integral formula. Thus, Eq. (\ref{61}) becomes
\ben
S(v,r)=-\int_{0}^{v}E(v')dv'+2\pi i\frac{4ME(v)}{N^{2}}.
\label{66}
\een

As a result, the wave function for the massless particle outgoing (and ingoing) may be written in the following forms:
\ben
\Psi_{out}(v,r)=exp\Big[\frac{i}{\hbar}\Big(-\int_{0}^{v}E(v')dv'+\pi i\frac{4ME(v)}{N^{2}}\Big)\Big],\nonumber\\
\label{67}
\een
\ben
\Psi_{in}(v,r)=exp\Big[\frac{i}{\hbar}\Big(-\int_{0}^{v}E(v')dv'-\pi i\frac{4ME(v)}{N^{2}}\Big)\Big].\nonumber\\
\label{68}
\een

For an outgoing particle, the tuneling rate is
\ben
\Gamma \sim e^{-2 \textit {Im}S} \sim e^{-2 \frac{4\pi ME(v)}{N^{2}}}= e^{-\frac{E(v)}{K_{B}T}}    
\label{69}
\een  
where $K_{B}=$ Boltzman Constant. 

Consequently
\ben
T_{H}=\frac{1}{8 \pi K_{B}}\frac{N^{2}}{M}=\frac{1}{8 \pi K_{B}}\frac{(1-\phi_{v}^{2})^{2}}{M},
\label{70}
\een
is the Hawking temperature. 

One can note that the usual Hawking temperature for the Vaidya spacetime~\cite{tang} is retrieved under the assumptions $\phi_{v}^{2}=0$ and $m(v,r)=M(v)$.\\

\section{Conclusion}
Manna et. al. \cite{gm1} have established a connection between the {\bf K-}essence geometry and the Vaidya spacetime, which produces a new spacetime called the {\bf K-}essence emergent Vaidya spacetime. In contrast to the usual Vaidya dynamical horizon equation, which is based on Sawayama's \cite{sawayama} finding, we have obtained the dynamical horizon equation (\ref{51}) for the spherically symmetric {\bf K-}essence emergent Schwarzschild-Vaidya spacetime. 

Some of the salient features of the present works are as follows:\\

(i) We have studied Hawking radiation in the {\bf K-}essence emergent Vaidya spacetime using the modified definition of the dynamical horizon~\cite{sawayama}. Using our work on Hawking radiation and applying the dynamical horizon equation, we get a transcendental equation (\ref{56}) that is far more difficult than the one found in Sawayama's~\cite{sawayama} transcendental equation (37). In the presence of the Wright omega function~\cite{corless1}, this transcendental equation (\ref{56}) can only be solved numerically, not analytically. 

(ii) Since $M\ne0$, the numerical solutions of the mass $m(v,r)$ always decrease with increasing $v$ but do not tend to zero. This is true for different values of $M$ and $r$, $\phi_{v}^{2}$. By analyzing the dynamical horizon equation as $\phi_{v}^{2}\rightarrow 0^{+}$, we have demonstrated analytically that the black hole mass $m(v,r)$ in the {\bf K-}essence emergent Schwarzschild-Vaidya spacetime constantly diminishes but does not completely evaporate. 

(iii) One key distinction between Vaidya spacetime and {\bf K-}essence emergent Schwarzschild-Vaidya spacetime is that the mass always disappears in Vaidya spacetime, but not in {\bf K-}essence emergent Schwarzschild spacetime. Nonetheless, we can witness the same trend that the black hole's mass shrinking in all scenarios. Possible explanations for the varied outcomes include the following: in our situation, standard gravity ($g_{\mu\nu}$) is minimally coupled with the scalar filed $\phi$, which causes additional interactions (forces) and the new form of emergent spacetime, as well as changes to the associated equation of motion and geodesic equation. 

(iv) We have calculated the Hawking temperature using the tunneling process~\cite{mitra1,siahaan}, which is distinct from the typical Vaidya case and the result is $T_{H}=\frac{1}{8 \pi K_{B}}\frac{(1-\phi_{v}^{2})^{2}}{M}$. In this work, we have used the {\bf K-}essence model as purely gravitational standpoint as~\cite{gm1,gm2,Ray}. 

(v) Finally, though the cosmic implications of dark energy are important in the current scenario, the {\bf K-}essence theory has received universal support as an explanation, and the $\phi_{v}^{2}$ quantity may be thought of as dark energy density in units of critical density~\cite{gm3,gm4,gm5}.\\

{\bf Acknowledgement:}
SR is thankful to the Inter-University Centre for Astronomy and
Astrophysics (IUCAA), Pune, India for providing Visiting Associateship
under which a part of this work was carried out who also gratefully
acknowledges the facilities under ICARD, Pune at CCASS, GLA
University, Mathura.\\

{\bf Data Availability Statement:} No Data associated in the manuscript.\\

\end{document}